\documentclass[pre,aps,twocolumn,floatfix,superscriptaddress,showpacs,nofootinbib]{revtex4-1}
\usepackage{amsmath,amssymb}
\usepackage{float,graphicx}
\usepackage{color}

\newcommand{\be}{\begin{equation}}
\newcommand{\ee}{\end{equation}}
\newcommand{\bea}{\begin{eqnarray}}
\newcommand{\eea}{\end{eqnarray}}
\newcommand{\bes}{\begin{subequations}}
\newcommand{\ees}{\end{subequations}}

\newcommand{\eref}[1]{Eq.~(\ref{#1})}%
\newcommand{\Eref}[1]{Equation~(\ref{#1})}%
\newcommand{\sref}[1]{Sec.~\ref{#1}}%

\begin{document}

\title{Stationary mass distribution and nonlocality in models of coalescence and shattering}
\author{Colm Connaughton}
\email{C.P.Connaughton@warwick.ac.uk} 
\affiliation {Mathematics Institute, University of Warwick, Gibbet Hill Road, Coventry CV4 7AL, UK}
\affiliation{Centre for Complexity Science, University of Warwick, Coventry CV4 7AL, UK}
\affiliation{London Mathematical Laboratory, 14 Buckingham St. London WC2N 6DF, UK}
\author{Arghya Dutta}
\email{argphy@gmail.com}
\altaffiliation[Present address: ]{Leibniz-Institut f\"ur Polymerforschung Dresden e. V.,
Institut Theorie der Polymere,
Hohe Stra\ss e 6, 01069 Dresden, Germany.}
\affiliation{Universit\'{e} de Strasbourg, CNRS, Institut Charles Sadron, UPR 22, 67000 Strasbourg, France}
\author{R. Rajesh}
\email{rrajesh@imsc.res.in}
\affiliation{The Institute of Mathematical Sciences, CIT Campus, Taramani, Chennai 600113, India}
\affiliation{Homi Bhabha National Institute, Training School Complex, Anushakti Nagar, Mumbai 400094, India}
\author{Nana Siddharth}
\email{nana.siddharth@gmail.com}
\affiliation{The Institute of Mathematical Sciences, CIT Campus, Taramani, Chennai 600113, India}
\affiliation{Homi Bhabha National Institute, Training School Complex, Anushakti Nagar, Mumbai 400094, India}
\author{Oleg Zaboronski}
\email{olegz@maths.warwick.ac.uk}
\affiliation {Mathematics Institute, University of Warwick, Gibbet Hill Road, Coventry CV4 7AL, UK}

\date{\today}
\begin{abstract}
We study the asymptotic properties of the steady state mass distribution for 
a class of collision kernels in an aggregation-shattering model in the limit of small shattering probabilities. 
It is shown that the exponents characterizing the large and small mass asymptotic behavior of the
mass distribution depend on whether the collision kernel is  local (the aggregation mass flux is essentially generated
by
collisions between particles of similar masses), or non-local (collision between particles of widely different masses give
the main contribution to the mass flux). We show that the non-local regime is
further divided into two sub-regimes corresponding to weak and strong non-locality. We
also observe that at the boundaries between the local and non-local regimes, the mass distribution acquires 
logarithmic corrections to scaling and calculate these corrections. Exact solutions
for special kernels and numerical simulations are used to validate some non-rigorous steps used in the analysis.
Our results show that for local kernels, the scaling solutions carry a constant flux of mass due to aggregation,
whereas for the non-local case there is a correction to the constant flux exponent. 
Our results suggest that for general scale-invariant kernels, the universality classes of mass distributions are labeled by
two parameters: the homogeneity degree of the kernel and one further number measuring the degree
of the non-locality of the kernel. 
\end{abstract}

\maketitle

\section{Introduction}

Many-body systems controlled by coalescence arise in many branches of science. The microscopic particles constituting such systems have a tendency to merge irreversibly upon collision or contact. 
Some examples include hydrogels for biomedical applications~\cite{Berger04}, supramolecular 
polymer gels~\cite{Sangeetha05}, aerosol formation~\cite{friedlander2000smoke}, cloud 
formation~\cite{falkovich2002,PNS2008}, ductile fracture~\cite{pineau2016failure}, and
charged biopolymers~\cite{tom2016aggregation,tom2017aggregation}.  By understanding the kinetics of coalescence, the macroscopic properties of such systems can be related to the microphysics of the fundamental collision and merging processes. For more applications and known results see the reviews~\cite{leyvraz_scaling_2003,handbook}. 
In some applications, colliding particles may also fragment or shatter into smaller particles. 
Whether the collision between particles will result in coagulation or fragmentation of the constituent particles 
depends on the energy of the colliding particles~\cite{wada2009collisional,brilliantov2009}. Typically particles with 
higher kinetic energy fragment, while slow moving ones coalesce or rebound. The size distribution of the fragmented
particles is typically a power law distribution~\cite{nakamura1991velocity,giblin1998properties,arakawa1999collisional,kun2006scaling}, reproducible
in simple tractable models~\cite{2014-snggb-njp-statistical,dhar2015fragmentation}. Such fragmentation processes
find application in geophysics~\cite{grady1980criteria}, astrophysics~\cite{michel2003disruption,nakamura2008impact},
glacier modeling~\cite{aastrom2014termini}, etc.

In this paper we are interested in situations where both coalescence and collisional fragmentation are simultaneously relevant. 
Examples include  the fluctuations of phase coherent domains in high-temperature superconductors \cite{johnson2011equivalent}, the statistical properties of insurgent conflicts \cite{bohorquez2009common},  the dynamics of herding behavior in financial markets \cite{teh2016asian}, and the formation and stability of planetary rings~\cite{davis1984,Longaretti1989} where a coalescence--fragmentation model has recently been proposed to explain the particle size distribution of Saturn's rings over several orders of magnitude \cite{brilliantov2009,brilliantov2015size}.  
When coalescence and fragmentation occur together, one might expect the system 
to reach a nonequilibrium steady state  in which the depletion of smaller particles due to coalescence is balanced by the depletion of larger particles due to fragmentation. 
These nonequilibrium states are expected to be insensitive to fine details of aggregation-fragmentation processes provided that the mass scales at which fragmentation acts as an effective source of light particles and the sink of heavy particles are widely separated.
The simplest model of fragmentation for which the described scale separation occurs naturally
is such that all fragmented particles are of the size of the smallest possible particle~\cite{brilliantov2009,brilliantov2015size}. 
We refer to such extreme fragmentation as shattering and use $m_0$ to denote the mass of the smallest particles generated.
The expected universality of  coalescence-shattering models explains the diversity of their applications and motivates a parametric study of how the particle size distribution depends on the form of the collision kernel, $K(m_1, m_2)$. The kernel
depends on the nature of particle motion and interaction and gives the dependence of the microscopic collision rate on the masses, $m_1$ and $m_2$ of the colliding particles.

 A particularly important class of collision kernels are homogeneous functions describing collisions that do not have a characteristic mass scale. 
This class includes many scientifically relevant cases, including the previously mentioned examples.  
We therefore restrict our attention to homogeneous kernels and denote the degree of homogeneity by $\beta$. 
Let us now consider how the stationary mass distribution should scale for a general homogeneous kernel.
In the limit of small shattering probability $p$, there is a divergent mass scale $M(p)$ beyond which there are very few heavy particles due to a high cumulative probability of shattering. 
For a large range of masses $m_0 \ll m \ll  M(p)$, we expect $N(m)$ to be such that the flux of mass due to coalescence $J(m)$ is $m$ independent due to local mass conservation. 
In the well-mixed limit, the mass scaling of the flux can be determined from the  following mean field scaling, (see \cite{connaughton2004stationary} for details):
\[
J(m) \sim m^3 m^\beta N^2(m).
\]
Constant aggregation flux, $J$,  implies that
\begin{eqnarray}\label{intro_KZ}
N(m)\sim m^{-\frac{3+\beta}{2}}.
\end{eqnarray}
By analogy with wave turbulence, we refer to the scaling exponent $(3+\beta)/2$ as the Kolmogorov-Zakharov   or constant flux exponent. 
It depends only on the kernel homogeneity, $\beta$, and therefore possesses a high degree of universality. 
In particular, it will not change if we perturb the kernel while preserving the value of $\beta$ or change the nature of 
of sinks and sources (e. g., by removing heavy particles from the system once they become heavier than a fixed mass $M$
as in \cite{connaughton2004stationary} or removing colliding particles at a certain rate as in \cite{Connaughton17}).

It is natural to ask when these constant flux solution is realized? 
This question can be answered by substituting the scaling solution as in Eq.~(\ref{intro_KZ}) into the analytic integral expression for the flux $J$ and checking that it remains finite in the limit of small shattering probability $p$ (see Refs.~\cite{connaughton2004stationary,crz2008} for details of 
the derivation). It turns out that the realizability of constant flux solution depends on the \textit{locality} of the collision mechanism. 
To characterize locality carefully, let us further reduce the class of kernels we are studying by assuming that in addition to having homogeneity degree $\beta$, $K(m_1, m_2)$ has the following asymptotic scaling when one of the colliding particles is much heavier than the other:
\[
K(m_1, m_2) \sim m_1^\mu m_2^\nu~\mbox{ for } m_2 \gg m_1.
\]
Clearly, $\beta = \mu + \nu$. This second reduction of generality is as natural for scale-free kernels as the homogeneity. It turns out that the realizabiliity of the Kolmogorov-Zakharov scaling depends, not on $\beta$, but on the difference
\[
\theta=|\mu-\nu|,
\]
which we call the locality exponent. In \cite{connaughton2004stationary}, we showed that for pure coalescence,  the constant flux scaling (\ref{intro_KZ})
is realized if the locality exponent $\theta<1$. 
Physically, kernels with $\theta<1$ lead to the flux of mass due to coalescence being dominated by collisions between similar mass particles. Hence the term ``locality.'' For pure coalescence, the scaling of the particle size distribution for local kernels is strongly universal: when the characteristic scales of the source and sink are widely separated,  it becomes independent of their details and depends only on the degree of homogeneity, $\beta$, of the collision kernel.  We therefore expect intuitively that for the coalescence-fragmentation case with $\theta < 1$, the scaling of the particle size distribution  in the limit of small shattering probability $p$ is given by the constant flux expression (\ref{intro_KZ}).

This  paper confirms this intuition and addresses the nonlocal case, $\theta >1$.  We will use
the simplest family of collision kernels labeled by two exponents $\nu$ and $\mu$ (or equivalently by $\beta=\nu+\mu$ and $\theta=\nu-\mu$):
\begin{eqnarray}\label{intro_ker_mod}
K(m_1,m_2) = m_1^\mu m_2^\nu + m_1^\nu m_2^\mu.
\end{eqnarray}
Without loss of generality, we assume that $\nu\geq \mu$. This family has been widely used in general studies of aggregation (see 
Refs.~\cite{leyvraz_scaling_2003,handbook} for  a review). 
Furthermore, we will always assume that the mean field approximation is applicable, which 
will allow us to calculate the mass distribution $N(m)$ by deriving and solving the corresponding
Smoluchowski equation~\cite{smol1917}.

For the family of kernels (\ref{intro_ker_mod}) we will show that if $\theta>1$, the scaling exponent of the mass distribution
in the limit of small shattering probability is both $\beta$- and $\theta$ dependent. Moreover, this dependence
is different depending on whether $1<\theta<2$ (weak non-locality) or $\theta>2$ (strong non-locality).
The amplitude of the mass distribution in the non-local regime is non-universal in the sense that it depends
on the effective shattering scale and the mass of dust particles. It is interesting to note that
these results for the aggregation-shattering model which conserves mass, parallel the answers
for a non-conserved system with coalescence, input of small particles,
and collision-dependent evaporation studied in ~\cite{Connaughton17}. It seems that
fine details of the mechanisms leading to effective sources and sinks of particles are irrelevant for a large class of coalescent models even in 
the non-local regime. An important conclusion from our analysis is that the mass distribution for the 
model kernel (\ref{intro_ker_mod}) for $\theta>1$ is different from the mass distribution for the 
local kernel $(m_1 m_2)^{\beta/2}$ with the same degree of homogeneity.

The paper is organized as follows. In \sref{sec:model} we define the model precisely and state our main 
quantitative
results. In \sref{sec:numerics} we discuss the numerical  algorithm that we  use to solve for the steady state mass 
distribution. It is an iterative procedure that we show to reproduce known exact solutions.  
In \sref{sec:exact} we solve the model exactly for two special cases: first when $\mu=\nu$ ($\theta=0$) and second   the addition model in which
two particles coalesce only when at least one particle is a monomer ($\theta=\infty$). These exact solutions help us
to benchmark the numerical algorithm. It is possible to obtain exact results for all integer $\theta$'s. This is 
discussed in \sref{sec:integer}, where the presence of  logarithmic corrections is established for some values
of $\theta$.
In \sref{sec:moments}, the small mass behavior of the mass distribution
is studied using the exact relations between different moments.  This  enables us to determine the exponents when the 
kernel is local, and relations between the exponents  when the kernel is non-local. 
In \sref{sec:singularity}, we analyze the large mass behavior of the mass distribution by studying the singularities of the 
generating functions. By stitching together the small and large mass behavior, we are able to determine both the
small and large mass asymptotic behavior of the mass distribution. In \sref{sec:rings} we discuss the implications of our findings for the specific case of planetary rings since it is an interesting example where both local and nonlocal cases may be relevant. Finally,
we conclude with a overview of results and directions of future research in \sref{sec:conclusion}.
\section{Model}
\label{sec:model}
Consider a collection of particles, each  characterized by a single scalar parameter, mass. The mass of particle $i$ will be
denoted by $m_i$, $i=1,2,,\ldots$, and  will be measured in terms of the smallest possible mass in the system $m_0$, corresponding
to the smallest possible dust particle, such that
$m_i$ is an integer. Given a certain initial configuration, the system evolves in time via coagulation and collision-dependent
fragmentation. Two particles of masses $m_1$  and $m_2$ collide at rate $(1+\lambda) K(m_1,m_2)$, where 
$K(m_1,m_2)$ is the  collision kernel. On collision, with probability $1/(1+\lambda)$, the two particles coalesce to
form a particle of mass $m_1+m_2$, and  with probability $\lambda/(1+\lambda)$, fragment into $(m_1+m_2)$ particles of
mass $1$. Note that both the dynamical processes conserve mass, so that total mass is a constant of motion. We would be
interested in the limiting case when the fragmentation rate tends to zero, i.e., $\lambda \to 0$. Also, we will be considering the 
well-mixed mean field limit when the spatial correlations between the particles may be neglected.

Let $N(m,t)$ denote the number of particles or mass $m$ per unit volume at time $t$. In the well-mixed dilute limit,
the time evolution of $N(m,t)$ is described by
the Smoluchowski equation:
\begin{widetext}
\begin{align}
\frac{d N(m,t)}{dt} =  &\frac{1}{2} \sum_{m_1=1}^{\infty}\sum_{m_2=1}^{\infty} 
N(m_1,t) N(m_2,t) K(m_1, m_2)\delta(m_1+m_2-m)- (1+\lambda) \sum_{m_1=1}^\infty N(m_1,t) N(m,t) K(m_1,m) \nonumber\\
&+ \frac{\lambda}{2}\delta_{m,1}\sum_{m_1=1}^{\infty}\sum_{m_2=1}^{\infty}N(m_1,t) N(m_2,t) K(m_1, m_2)(m_1+m_2). 
\label{eq:model}
\end{align} 
\end{widetext}
The first term in the right hand side of \eref{eq:model} is a gain term that accounts for the number of ways a particle of mass $m$ 
may be created through a coalescence event. The second term is a loss term that accounts for the number of ways in which
$N(m,t)$ decreases due to  coalescence or fragmentation.  The last term describes the creation of particles of mass 
$1$ due to  fragmentation events. It is easy to check that the mean mass density is conserved. 
In this paper, we will be interested in the steady state solution of \eref{eq:model} obtained by setting the time derivative
to $0$. We will denote the steady state solution by $N(m)$. $N(m)$ satisfies the equation
\begin{widetext}
\begin{align}
0=&\frac{1}{2} \sum_{m_1=1}^{\infty}\sum_{m_2=1}^{\infty} 
N(m_1) N(m_2) K(m_1, m_2)\delta(m_1+m_2-m)- (1+\lambda) \sum_{m_1=1}^\infty N(m_1) N(m) K(m_1,m) \nonumber\\
&+ \frac{\lambda}{2}\delta_{m,1}\sum_{m_1=1}^{\infty}\sum_{m_2=1}^{\infty}N(m_1) N(m_2) K(m_1, m_2)(m_1+m_2). 
\label{eq:model1}
\end{align} 
\end{widetext}

We consider the  general class of kernels given by
\begin{align}
K(m_1,m_2) = m_1^\mu m_2^\nu + m_1^\nu m_2^\mu,\quad \nu \geq \mu.
\label{eq:kernel}
\end{align}
The kernel may also be classified using two other exponents.  The first is the homogeneity exponent $\beta$ defined through
$K(hm_1, hm_2)=h^\beta K(m_1, m_2)$:
\be
\beta=\nu+\mu.
\ee
The second is the nonlocality exponent $\theta$ defined as
\be
\theta=\nu-\mu.
\ee
When $\beta>1$, the kernel is referred to as a gelling kernel and non-gelling otherwise. We will refer to 
kernels with $\theta<1$ as local kernels and non-local otherwise.

We also consider another kernel that corresponds to the so called addition model~\cite{Hendriks1984,brilliantov1991,laurencot1999,ball_instantaneous_2011,blackman1994coagulation,chavez1997some} . 
Here collision events are allowed
only if at least one of the particles has mass $1$. The kernel for the addition model is
\begin{align}
\label{eq:kernel-addition}
K^{add}(m_1,m_2)=m_1^\nu m_2^\nu (\delta_{m_1,1}+\delta_{m_2,1}),
\end{align}
which is characterized by a single exponent $\nu$. This
kernel turns out to be exactly solvable (see \sref{sec:addition}).

In this paper, we will determine the asymptotic behavior of $N(m)$ through analysis of the
moments as well as the singularities. 
Moments and generating function are defined as:
\begin{align}
\label{eq:mom-def}
{\mathcal M}_\alpha &=  \sum_{m=1}^\infty m^\alpha N(m) ,\\
F_\alpha (x) &= \sum_{m=1}^\infty m^\alpha N(m) x^m.
\end{align}
Clearly $F_\alpha(1)=\mathcal{M}_\alpha$.  Multiplying \eref{eq:model1} by $x^m$ and summing over all $m$, 
we obtain a relation between moments and generating functions,
\begin{align}
\label{eq:gen-fun}
&F_\mu (x) F_\nu (x)-(1+\lambda)\left[{\mathcal M}_\mu F_\nu (x)+{\mathcal M}_\nu F_\mu (x)\right] \nonumber\\
&+x(1+2\lambda){\mathcal M}_\mu{\mathcal M}_\nu=0.
\end{align}
\begin{table}
\caption{\label{table1} Summary of results obtained in this paper. 
The exponents $y$, $\tau_s$, $\eta_s$, $\tau_\ell$,
and $\eta_\ell$ are as defined in Eqs.~\eqref{ydefn}, \eqref{eq:smallmass},
and \eqref{eq:exponential}.  For $\theta=1$ and $2$, there are 
additional logarithmic corrections as described in Eqs.~\eqref{eq:borderline} and \eqref{eq:n=2},
respectively.
}
\begin{ruledtabular}
\begin{tabular}{rccccc}
$\theta$&$y$& $\tau_s$ & $\eta_s$& $\tau_\ell$ & $\eta_\ell$\\
\hline
$0$ & $2$ & $\frac{3 +\beta}{2}$ &  $\max[0,\frac{1-\beta}{2}]$ & $\frac{3 +\beta}{2}$ &$\max[0,\frac{1-\beta}{2}]$ \\
$(0,1)$& $\frac{2}{\theta+1}$ & $\frac{3 +\beta}{2}$ & $\max[\frac{1-\beta}{2},0]$ & $\frac{2 +\beta}{2}$ & $  \max[\frac{2-\beta}{2},\frac{1}{2}]$ \\
$ (1,2)$ & $1$ & $\mu+2$ & $\max[-\mu, 0 ]$ & $\frac{2 +\beta}{2}$ & $\eta_s+\frac{2-\theta}{2}$ \\
$ (2,\infty)$ & $1$ & $ \nu$ & $\max[2-\nu,0]$ & $\nu$ & $\max[2-\nu,0]$ 
\end{tabular}
\end{ruledtabular}
\end{table}

We also define the exponents that characterize the mass distribution $N(m)$.
We assume that  the only relevant mass scale in the problem is the cutoff mass $M$ and hence
$N(m)$ has the scaling form:
\begin{align}
\label{eq:scaling-main}
N(m)=m^{-\tau}f\left(\frac{m}{M}\right),\; m,M\gg1,
\end{align}
where $\tau$ is an exponent and $f(x)$ is a scaling function. $M$ denotes the cutoff scale below and above which $N(m)$
behaves differently. There are two cutoff mass scales in the problem. One is the total mass in the system and the
other is the scale introduced by fragmentation. We will be working in the limit when total mass is infinite, but mean density
is finite, leaving only one cutoff scale. The divergence of the cutoff mass scale as the fragmentation rate $\lambda \to 0$ is captured by
\be
M\sim \lambda^{-y}, \quad \lambda \to  0,
\label{ydefn}
\ee
where the exponent $y$ will depend on the kernel.
To characterize the  scaling behavior for small and large masses,  we introduce four new exponents 
$\tau_s$, $\eta_s$, $\tau_{\ell}$ and $\eta_{\ell}$ which are defined as:
\begin{align}
N(m) \simeq& \frac{a_s}{m^{\tau_s} M^{\eta_s}},~ m \ll M,
\label{eq:smallmass}\\
N(m) \simeq& \frac{a_{\ell} }{m^{\tau_\ell} M^{\eta_\ell}} e^{-m/M} ,~ m \gg M.
\label{eq:exponential}
\end{align}
The exponential decay for large mass is a conjecture. For $m \gg M$, the exponential decay with mass will be  supported by 
exact solutions for special cases  and numerical observation for more general cases.  Further justification
for arbitrary kernels follows from the additivity principle using which it has been 
argued that, for generic  conserved mass models, the mass distribution
has an exponential decay~\cite{cpm2014,dcp2016}.
The four exponents are not independent. It is straightforward to obtain from
Eq.~(\ref{eq:scaling-main}) that
\begin{equation}
\tau_s+ \eta_s=\tau_\ell + \eta_\ell = \tau.
\label{eq:expequality}
\end{equation}
The results obtained in this paper for the different exponent as a function of the
exponents $\theta$ and $\beta$ are summarized in Table.~\ref{table1}.

\section{Numerical Algorithm \label{sec:numerics}}

In this section, we describe the numerical scheme for obtaining the steady state
mass distribution $N(m)$. Solving \eref{eq:model1}  in the
steady state for $m=1$, we obtain 
\be
N(1) = \frac{2 \lambda+1}{1+\lambda} \frac{{\mathcal M}_\mu {\mathcal M}_\nu}{{\mathcal M}_\mu +{\mathcal M}_\nu}.
\label{eq:n1}
\ee
For $m \geq 2$, $N(m)$ may be 
determined from \eref{eq:model1} in the steady state, provided 
${\mathcal M}_\mu$,  ${\mathcal M}_\nu$,
and all $N(k)$ for $k<m$ are known:
\be
N(m) = \frac{\displaystyle\sum_{m_1=1}^{m-1} N(m_1) N(m-m_1) K(m_1,m-m_1) }{2(1+\lambda)(m^\mu {\mathcal M}_\nu + m^\nu
 {\mathcal M}_\mu)}.
 \label{eq:nm}
\ee
Thus, ${\mathcal
M}_\mu$ and ${\mathcal M}_\nu$ determine $N(m)$ for $m\geq 1$.

Consider scaled variables $\widetilde{N}_m=N(m)/N(1)$ and $\widetilde{{\mathcal M}}_\alpha= {\mathcal M}_\alpha/N(1)$.
In terms of these variables, Eqs.~(\ref{eq:n1}) and (\ref{eq:nm}) reduce to
\bea
\label{eq:n1scaled}
1& =& \frac{2 \lambda+1}{1+\lambda} \frac{\widetilde{{\mathcal M}}_\mu \widetilde{{\mathcal M}_\nu}}{\widetilde{{\mathcal M}}_\mu +\widetilde{{\mathcal M}}_\nu},\\
 \label{eq:nmscaled}
\widetilde{N}(m) &=& \frac{\displaystyle\sum_{m_1=1}^{m-1} \widetilde{N}(m_1) \widetilde{N}(m-m_1) K(m_1,m-m_1) }
{2(1+\lambda)(m^\mu \widetilde{{\mathcal M}}_\nu + m^\nu
 \widetilde{{\mathcal M}}_\mu)}.
\eea
$\widetilde{{\mathcal
M}}_\mu$ and $\widetilde{{\mathcal M}}_\nu$ determine $\widetilde{N}(m)$ for $m\geq 1$.
The two unknowns, $\widetilde{{\mathcal M}}_\mu$ and $\widetilde{{\mathcal M}}_\nu$ are 
not independent and related
to each other through  \eref{eq:n1scaled}.

\begin{figure} 
\includegraphics[width=\columnwidth]{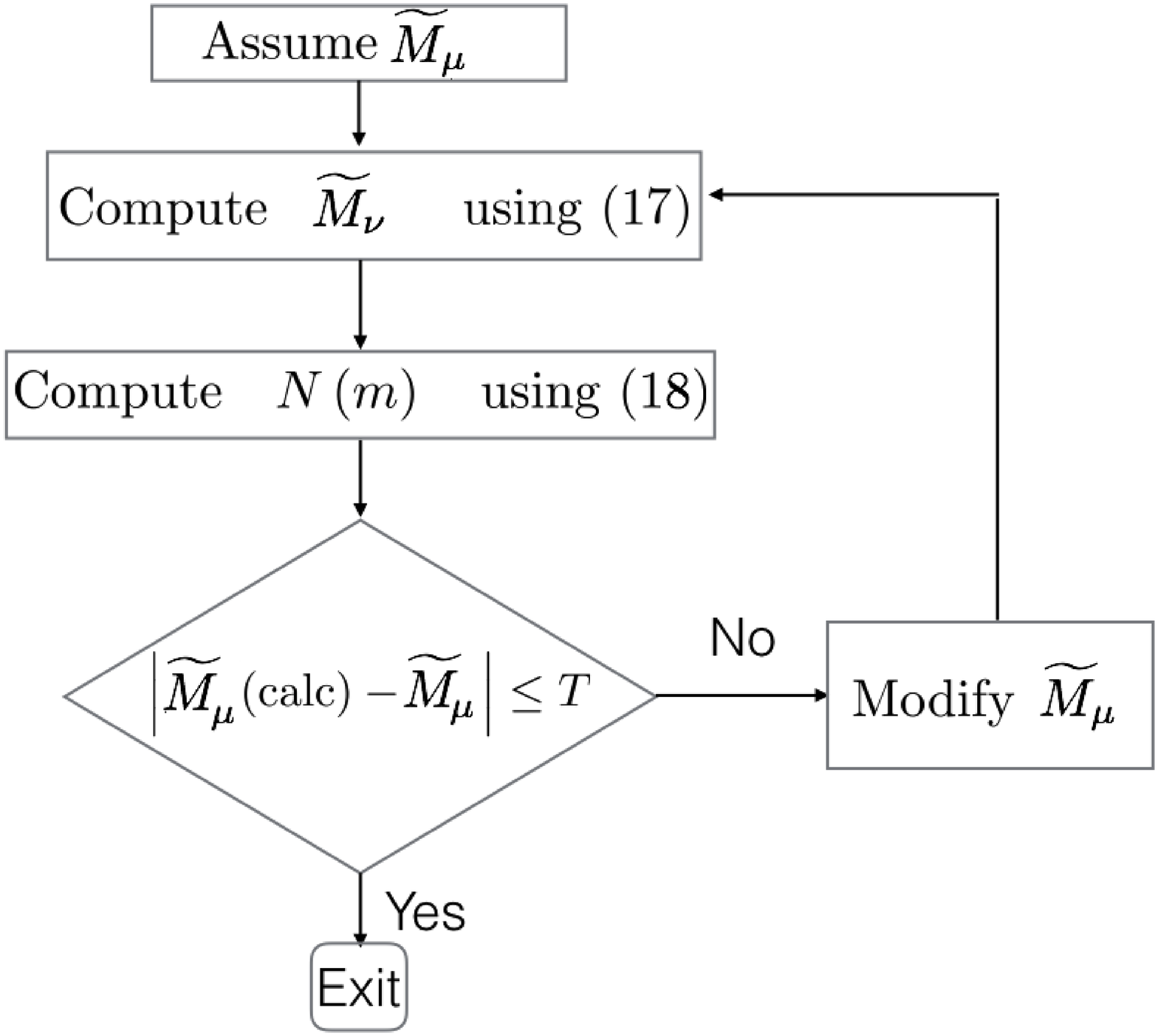} 
\caption{Flowchart describing the iterative numerical algorithm for determining the
steady state distribution $N(m)$.
} 
\label{fig:flowchart}
\end{figure}

To determine $\widetilde{{\mathcal M}}_\mu$, we follow an iterative procedure as
summarized in the flowchart shown in Fig.~\ref{fig:flowchart}.
We start by assigning a numerical value (close to $1.0$)  for $\widetilde{{\mathcal M}}_\mu$.
$\widetilde{{\mathcal M}}_\nu$ is determined from \eref{eq:n1scaled}. We then
solve for $\widetilde{N}(m)$ up to a value of 
$m$ for which $\widetilde{N}(m)$ is larger than a predetermined value
($10^{-16}$ in our analysis), setting $N(m)=0$ for larger values of $m$. We then check for
self consistency, i.e whether $\sum m^\mu \widetilde{N}(m)$ is equal to the preassigned
value of $\widetilde{{\mathcal M}}_\mu$. 
We increment $\widetilde{{\mathcal M}}_\mu$ in small steps until the self-consistency 
condition is satisfied to the required precision. 
In our numerical analysis, we demand that the
difference between the assumed and calculated values of 
$\widetilde{{\mathcal M}}_\mu$ should be smaller than $10^{-10}$. 

To determine the unscaled variables $N(m)$, we use the fact that mass is 
conserved: $\sum_1^\infty m N(m) = \rho$, where $\rho$ will be treated as a parameter.
We then scale all $\widetilde{N}(m)$ by the same factor so that the desired mass density
$\rho$ is achieved, thereby determining $N(m)$. In our numerical measurements, we set $\rho=1$.
There is no proof that the algorithm will result in the convergence of $N(m)$ to its correct value.
However, we verify the convergence for special solvable kernels (see Sec.~\ref{sec:exact}), leading
us to expect that the mass distribution converges to its correct value for more general kernels.

From the numerically computed $N(m)$, we observe that, for all values of 
$\mu$ and $\nu$ that we have studied, $N(m)$ decays exponentially at large masses [as in
\eref{eq:exponential}]. The exponential cutoff mass $M$ is determined by solving
for the three parameters ($\tau_\ell$, $M$, and $a_\ell/M^{\eta_\ell}$) in Eq.~(\ref{eq:exponential}) using 
$N(m)$ for three consecutive
$m$'s and extrapolating to large $m$. Once $M$ is determined, 
the compensated mass distribution
$N(m) e^{m/M}$ is a power
law with exponent $\tau_\ell$, allowing us to verify the theoretical
results for the exponents at large mass.

\section{Exact solutions \label{sec:exact}}
The steady state mass distribution $N(m)$ may be determined exactly for two cases:  (1) when $\mu=\nu=\beta/2$ and (2) the
addition model (defined in \sref{sec:addition}). 

\subsection{ Multiplicative kernel: $\mu=\nu=\beta/2$ ($\theta=0$)}

When $\mu=\nu=\beta/2$, the kernel \eref{eq:kernel} reduces to the multiplicative kernel. In this case,  \eref{eq:gen-fun} for
the generating function reduces to the quadratic equation
\be
F_{\beta/2}^2(x) -2 (1+\lambda) {\mathcal M}_{\beta/2} F_{\beta/2} (x) 
+x(1+2\lambda){\mathcal M}_{\beta/2}^2=0
\ee
which may be solved to yield
\be
\label{eq:fmu}
F_{\beta/2}(x)=(1+\lambda){\mathcal M}_{\beta/2}\left[ 1-\sqrt{1-\frac{x}{x_c}}\right],
\ee
where
\be
\label{eq:x0}
x_c=\frac{(1+\lambda)^2}{(1+2\lambda)},
\ee
and the sign of the square root of the discriminant is fixed by the constraint $F_\mu(0)=0$.
 The coefficient of $x^m$ is the Taylor expansion of $F_\mu(x)$ is $N(m)$ and is:
\begin{align}
\label{eq:nm-exact}
N(m)=\frac{(2m-2)!}{2^{2m-2}m!(m-1)!}\frac{N(1)}{m^{\beta/2} x_c^{m-1}}, ~m=2, 3,\ldots,
\end{align}
where
\begin{align}
\label{eq:n1-exact}
N(1)=\frac{{\mathcal M}_{\beta/2}(1+\lambda)}{2 x_c}.
\end{align}
For large $m$, the factorials may be approximated using 
Stirling formula, and the asymptotic behavior of $N(m)$ for large $m$ may be derived to be
\begin{align}
\label{eq:nm-asymp}
N(m)\simeq\frac{N(1) x_c}{\sqrt{\pi}}\frac{e^{-m/M}}{m^{(3+\beta)/2}},~m\gg 1,
\end{align}
where 
\be
M=\frac{1}{\lambda^2},~\lambda \to 0, 
\ee
or equivalently the exponent $y=2$. 

In \eref{eq:nm-asymp}, $N(1)$ is determined by the condition that ${\mathcal M}_1 =\rho$ is a constant. It is
not possible to find a closed form expression for $N(1)$ for arbitrary $\beta$, however, when $\beta/2$
is an integer, it is possible to determine it by differentiating or integrating \eref{eq:fmu} with respect to $x$ and
setting $x=1$. It is then straightforward to obtain $N(1)=\frac{2\lambda (1+\lambda)\rho}{2 x_c(1+2\lambda)}$ for $\beta=0$, and
$N(1)=\frac{\rho (1+\lambda)}{2 x_c}$ for $\beta=2$. For generic $\beta$, we use the asymptotic form \eref{eq:nm-asymp} to obtain
the dependence of $\langle m \rangle$ on the cutoff, thus determining $N(1)$. We thus obtain
\be
\label{eq:exact-main} 
N(m)\propto \frac{\rho}{M^{\max[0,(1-\beta)/2]}}
\frac{e^{-m\lambda^2}}{m^{(3+\beta)/2}}, \quad \theta=0.
\ee
In the limit of $\lambda \to 0$, $N(m)$ tends to a finite limit only when the kernel is gelling, i.e., $\beta>1$. For non-gelling
kernels with $\beta<1$, the prefactor tends to zero with decreasing fragmentation rate $\lambda$. This observation
may be
rationalized by the
fact the mass capacity of gelling kernels is finite and infinite for non-gelling kernels.
Summarizing the results for $\theta=0$, we have derived the results
\bes
\label{theta0-results}
\begin{align}
&\tau_s=\tau_\ell=\frac{3+\beta}{2},\\
&\eta_s=\eta_\ell= \max[0,(1-\beta)/2],\\
&y=2.
\end{align}
\ees

\begin{figure} 
\includegraphics[width=\columnwidth]{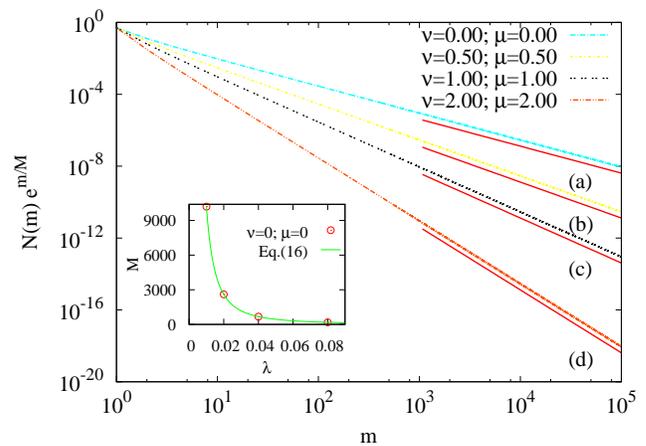} 
\caption{$N(m) e^{m/M}$ when $\nu=\mu$ for $\nu=0, 1/2,1,2$. The
solid lines are power laws with an exponent $-(3+\beta)/2$: (a) $-3/2$,
(b) $-2$, (c) $-5/2$, and (d) $-7/2$. The evaporation rate is $\lambda=0.01$. 
Inset: $M$, obtained from numerical analysis, is compared with 
the analytical result in Eq.~(\ref{eq:x0}). 
} 
\label{fig_nueqmu}
\end{figure}

The exact solution for the case $\mu=\nu$  
can be used to benchmark the numerical scheme described in Sec. \ref{sec:numerics}. 
In Fig.~\ref{fig_nueqmu}, we plot the numerical solution to 
Eq.~(\ref{eq:model1}), obtained using the aforementioned algorithm,
for four
different values of $\mu$.  The numerically determined cutoff scale $M$ 
is in excellent agreement with the exact solution 
(see inset of Fig.~\ref{fig_nueqmu}). The data for the compensated
mass distribution $N(m) e^{m/M}$ are power laws with exponents 
matching the ones obtained from the exact solution 
(see Fig.~\ref{fig_nueqmu}). We thus conclude that
the numerical scheme is accurate and stable. 


\subsection{Addition model with fragmentation \label{sec:addition}}
In this section, we calculate $N(m)$ for the addition model. In this case, collisions between particles are allowed only
if at least one of the masses is one, and the resulting collision kernel is as in \eref{eq:kernel-addition}. While this model is expected
to mimic the kernel \eref{eq:kernel} with $\nu \gg \mu$ when collisions between dissimilar masses dominate,
the exact regime of applicability will become clear only on comparing with the full solution for $N(m)$.
For the addition model, the Smoluchowski equation \eref{eq:model1}  in the steady state is
\begin{align}
&0= N(1)\left[(m-1)^\nu N(m-1)-(1+\lambda)m^\nu N(m)\right]\nonumber\\
&+\delta_{m,1} \left[\lambda N(1)(N_\nu+N_{\nu+1})-(1+\lambda)N_\nu N(m) m^\nu \right].
\end{align}
This is easily solved to give
\be
\label{eq:n1-add}
N(m)=\frac{\lambda {\mathcal M}_{\nu+1}-{\mathcal M}_{\nu}}{m^\nu(1+\lambda)^{m}}, ~m=1, 2,\ldots,
\ee
which in the limit of large mass and vanishing $\lambda$ reduces to
\be
N(m)\simeq \frac{N(1)}{m^\nu}e^{-m/M},\quad m \gg 1,
\ee
where $M=\lambda^{-1}[1+O(\lambda)]$. The unknown parameter $N(1)$ is determined by the constraint
that mass is conserved: $\sum_1^\infty mN(m) =\rho$. We thus obtain
\be
\label{eq:addition-main}
N(m)\propto \frac{\rho}{M^{\max[2-\nu,0]}}\frac{e^{-m/M}}{m^\nu},~~m \gg 1.
\ee
To summarize, we have obtained
\bes
\label{addition-results}
\begin{align}
&\tau_s=\tau_\ell=\nu,\\
&\eta_s=\eta_\ell= \max[0,2-\nu],\\
&y=1,
\end{align}
\ees
for the addition model.
\section{Mass distribution for integer $\theta$ \label{sec:integer}}

It is possible to obtain exact results for the case when the locality exponent $\theta$ is an integer,  namely $\theta=n$, $n$ is an integer.
The starting point is Eq.~\eqref{eq:gen-fun}. If $\nu=\mu+n$, where $n=1,2,\ldots$,
Eq.~\eqref{eq:gen-fun} reduces to a closed differential equation for $F_\mu$:
\begin{align}\label{eq:eq1}
&[F_\mu(x)-(1+\lambda){\mathcal M}_\mu] (x\partial_x)^nF_\mu(x)-(1+\lambda) {\mathcal M}_\nu F_\mu(x)\nonumber\\
&+x (1+2\lambda){\mathcal M}_\mu{\mathcal M}_\nu =0.
\end{align}
We expect the singularities to occur at the points in the complex $x$-plane 
where the coefficient in front of the highest order term is zero.
Therefore, at  the singular point $x_c$, $F_\mu$ satisfies 
\begin{eqnarray}
F_{\mu}(x_c)=(1+\lambda){\mathcal M}_\mu.
\end{eqnarray}

Introduce new variables $f(x)$ as follows:
\begin{eqnarray}
F_\mu(x) &=& (1+\lambda){\mathcal M}_\mu+f(x),\\
t &=&\ln(x),
\end{eqnarray}
where $f(x_c)=0$ and $t_c=\ln(x_c)>0$.
Then Eq.~\eqref{eq:eq1} may be rewritten as
\be
\label{eq:fg}
f \partial_t^n f -(1+\lambda) {\mathcal M}_\nu f-(1+2\lambda){\mathcal M}_\mu{\mathcal M}_\nu\left[\frac{(1+\lambda)^2}{(1+2\lambda)}-e^t\right]=0.
\ee
Equation~\eqref{eq:fg} becomes more tractable under the following transformations:
\bea
t &= &\ln\left(\frac{(1+\lambda)^2}{(1+2\lambda)}\right)+\tau,\\~
f(t) &=& (1+\lambda){\mathcal M}_{\nu}g(\tau),
\eea
such that we obtain
\be
\label{me}
g(\tau)\partial_\tau^n g(\tau)-g(\tau)-j(1-e^\tau)=0,
\ee
where
\be
j=\frac{{\mathcal M}_\mu}{{\mathcal M}_\nu}.
\ee
We note that $g(\tau_c)=0$. We now analyze Eq.~\eqref{me} for specific integer values of
$\theta$.

\subsection{$\theta=1$}
When $n=1$, near the critical point Eq.~\eqref{me} reduces to
\be
gg'=j(1-e^{\tau_c})+o(1),
\ee
since $g(\tau_c)=0$. Solving for $g(\tau)$, we obtain
\be
g(\tau)=\sqrt{2j(e^{\tau_c}-1)}\sqrt{\tau_c-\tau}+o(\sqrt{\tau_c-\tau}).
\ee
The generating function $g$ must be real for $\tau \in \mathbf{R}$, $\tau<\tau_c$.
Therefore we must have $\tau_c>0$ or in terms of the original variables, 
\be
x_c>\frac{(1+\lambda)^2}{(2\lambda+1)}.
\ee
We conclude that for $\theta=1$,
\be
f(x)=A\sqrt{x_c-x}+o(\sqrt{x_c-x}),
\ee
where
the amplitude is
\be
A=\sqrt{2(1+2\lambda){\mathcal M}_\mu{\mathcal M}_\nu \left[1-\frac{(1+\lambda)^2}{(1+2\lambda)}x_c^{-1}\right]}.
\ee
We conclude that in the limit of large masses,
\bea
N_\mu(m)&\sim & A\int_{0}^\infty \frac{dx}{\pi} (x_c+x)^{-m-1} \sqrt{x}\\ 
&\sim &\frac{A x_c^{1/2}}{2\sqrt{\pi}} m^{-3/2}e^{-m\ln(x_c)}.
\eea
Equivalently,
\be\label{ans_1}
N(m)\sim \sqrt{\frac{(1+2\lambda){\mathcal M}_\mu{\mathcal M}_\nu}{2\pi} \!\! \left[x_c-\frac{(1+\lambda)^2}{(1+2\lambda)}\right]} 
\frac{x_c^{-m}}{m^{\mu +3/2}}.
\ee

An independent moment equations analysis shows that for $\lambda \downarrow 0$  [see 
Eqs.~\eqref{eq:y-non-local} and \eqref{eq:Mx0}],
\be
x_c=1+\frac{1}{M},\mbox{ where } M\sim \lambda^{-1}.
\ee
Then the small $\lambda$ limit of Eq.~\eqref{ans_1} is
\begin{equation}
N(m)\sim \sqrt{\frac{{\mathcal M}_\mu{\mathcal M}_\nu}{2\pi M}} \frac{e^{-m/M}}{m^{(2+\beta)/2}}, ~\theta=1.
\end{equation}

\subsection{$\theta=2$}
When $n=2$, near the critical point Eq.~(\ref{me}) reduces to
\be
gg''=j(1-e^{\tau_c})+o(1), ~~g(\tau_c)=0,
\ee
which has a solution given by
\[
g(\tau)=\sqrt{2j(e^{\tau_c}-1)\ln\frac{\Delta}{\tau_c-\tau}} (\tau_c-\tau)+\ldots,
\]
where $\Delta$ is a positive constant which sets a reference scale in $\tau$-space. Notice that 
the solution depends on an arbitrary constant $\Delta$, which is consistent with the  solution of 
a second order ordinary differential equation subject to a single boundary condition $g(\tau_c)=0$. 
In principle, $\Delta$ may be determined by matching this singularity dominated solution with
the solution far from the singular point.
In the original variables this reads as
\begin{eqnarray}
f(y)=\sqrt{2J\left[x_c-\frac{(1+\lambda)^2}{(1+2\lambda)}\right]}\frac{y}{x_c}\sqrt{\ln\frac{\Delta  x_c}{y}}+\ldots,
\end{eqnarray}
where $y=x_c-x$, and
\be
J=(1+2\lambda){\mathcal M}_\mu{\mathcal M}_\nu.
\ee

Calculating the jump over the branch cut singularity of $f$, we find that
\be
N_\mu(m) =\sqrt{\frac{J}{2}\left[x_c-\frac{(1+\lambda)^2}{1+2\lambda}\right]}
 \int_0^\infty \!\!\!\!\!dy 
\frac{y [\ln\frac{\Delta  x_c}{y}]^{-1/2}}{x_c(x_c+y)^{m+1}}+\ldots 
\ee
Changing variables $y\rightarrow yx_c/m$ and taking the large-$m$ limit of the integral we obtain
\be
N(m)\sim\sqrt{\frac{J}{2}\left[x_c-\frac{(1+\lambda)^2}{(1+2\lambda)}\right]}
\frac{e^{-m\ln x_c}}{m^{\nu}\sqrt{\ln\frac{m}{m_0}}},
\label{eq:79}
\ee
where $m_0$ is a reference scale in the mass space.

In the limit of small $\lambda$, we expect that [see Eqs.~\eqref{eq:y-non-local} and \eqref{eq:Mx0}]
\be
x_c=1+\frac{1}{M},\mbox{ where } M\sim \lambda^{-1}.
\ee
Then, Eq.~\eqref{eq:79} simplifies to
\begin{eqnarray}
\label{eq:n=2}
N(m)\sim\sqrt{\frac{{\mathcal M}_\mu{\mathcal M}_\nu}{2M}}
\frac{e^{-m/M}}{m^{\nu}\sqrt{\ln\frac{m}{m_0}}},
\end{eqnarray}
Therefore, we find that there are logarithmic corrections to the scaling form, and $\theta=2$, $\tau_\ell=\nu$, $\eta_\ell=1/2$ and $a_\ell=\sqrt{J/2}$.


\subsection{$\theta=3,4,\ldots$}
Finally, we analyze Eq.~(\ref{me}) for $n>2$. Near the critical point,
\be
\label{men}
g(\tau)\partial_\tau^n g(\tau)=j(1-e^{\tau_c})+\ldots.
\ee
We try the  family of solutions:
\begin{equation}\label{solnn}
g(\tau)=p_{n-2}(\tau_c-\tau)+A(\tau_c-\tau)^{n-1}\log\left(\frac{\Delta}{\tau_c-\tau}\right)+\ldots,
\end{equation}
where $p_{n-2}$ is a polynomial of $(n-1)$-st degree such that $p_{n-2}(0)=0$,
\be
p_n(x)=d_1x+d_2x^2+\ldots+d_{n-2}x^{n-2}.
\ee
Note that Eq.~\eqref{solnn} depends on $n$ arbitrary constants [before we impose the condition $g(\tau_c)=0$],
which makes it a good candidate for a general solution.
Differentiating the above ansatz, we find:
\be
\partial_\tau^n g(\tau)=(-1)^n n! \frac{A}{\tau_c-\tau}
\ee
Substituting this into Eq.~\eqref{men} gives an answer for the amplitude:
\be
A=\frac{(-1)^{n-1}}{(n-1)!}\cdot \frac{j(1-e^{\tau_c})}{d_1}.
\ee
The coefficient $d_1$ can be expressed in terms of $F_{\mu+1}$: it follows from the definition of $F_\mu$ that
\be
d_1=\partial_\tau g(x_c)=\frac{F_{\mu+1}(x_c)}{(1+\lambda) {\mathcal M}_\nu}.
\ee

Applying the inversion formula and using the fact that the analytic part of $g(\tau)$ does not
contribute to the large mass asymptotic, one finds that
\begin{equation}
N(m)\sim \frac{J}{F_{\mu+1}(x_c)}\left[x_c-\frac{(1+\lambda)^2}{1+2\lambda}\right] \frac{e^{-m\ln(x_c)}}{m^\nu}.
\end{equation}
For small $\lambda$'s,
\begin{equation}
N(m)\sim \frac{{\mathcal M}_\mu{\mathcal M}_\nu}{MF_{\mu+1}(x_c)} \frac{e^{-m/M}}{m^\nu}, ~m\gg M, ~\theta=3,4,\ldots
\end{equation} 

\begin{figure}
\includegraphics[width=\columnwidth]{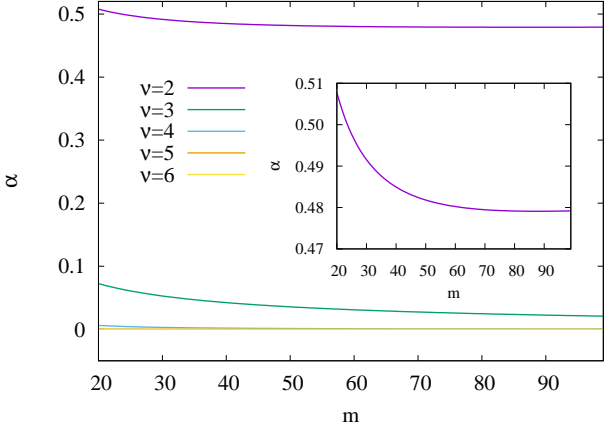}
\caption{Variation of the  exponent $\alpha$ characterizing the logarithmic corrections [see
Eq.~\eqref{eq:log_sum}] with $m$ for different values of $\nu$. The data are for $\mu=0$,
$\lambda=20.0$, and different $\nu$. The exponent $\alpha$ converges to $\alpha=0 (\nu\neq 2)$, and $\alpha=1/2 (\nu =2)$. 
Inset: The variation $\alpha$ with $m$ for $\nu=2$
is shown separately for clarity. }
\label{fig:logcorrections}
\end{figure}

The logarithmic corrections to the mass distribution for integer $\theta$, calculated in this section may be 
summarized as follows: 
\be
N(m)\approx \frac{a_\ell e^{-m/M}}{m^{\nu}(\ln m)^{\alpha}},~m\gg M, ~\theta=2,3,\ldots,
\label{eq:log_sum}
\ee
where $\alpha=1/2$ for $\theta=2$ and zero otherwise. 
We also note that these results coincide with the results obtained using analysis of singularities for non-integer
$\theta>2$ [see Eq.~\eqref{eq:61}].
The solution \eqref{eq:log_sum} is now verified using the numerical solution for $N(m)$ for integer $\theta$. 
Equation~\eqref{eq:log_sum} has three unknown
parameters $a_\ell$, $M$, and $\alpha$.
These parameters are determined as a function of $m$ by using $N(m)$ for three consecutive $m$.  The variation of $\alpha$
with $m$ is shown in Fig.~\ref{fig:logcorrections}. In these data, a large value of $\lambda$ ($\lambda=20.0$) is
chosen so that the small mass regime is suppressed and 
the large mass regime is exaggerated. It can be seen from Fig.~\ref{fig:logcorrections} that the exponents converge, albeit slowly, to their
predicted theoretical values [see Eq.~\eqref{eq:log_sum}].

\section{Moment analysis \label{sec:moments}}
An exact solution is possible only when $\theta=0$. In this section, we  use moment analysis to
determine some of the exponents characterizing $N(m)$ for general $\theta$ and $\beta$. In particular, we
study the small mass behavior of the mass
distribution $N(m)$ as described by Eq.~\eqref{eq:smallmass}.
Our aim is to determine the exponents $\tau_s$, $\eta_s$, and $y$
as a function of $\beta$ and $\theta$.  For this, we will 
require the equations satisfied by the different moments of $m$.
These may be obtained by
differentiating \eref{eq:gen-fun} with respect to $x$ or by multiplying
\eref{eq:model1} by $m^n$ and summing over $m$ from $1$ to $\infty$. Doing so gives
\begin{align}
\label{eq:high-mom}
& \lambda ({\mathcal M}_{\mu+n}{\mathcal M}_\nu+{\mathcal M}_\mu{\mathcal M}_{\nu+n})=
(1+2\lambda){\mathcal M}_\mu{\mathcal M}_\nu\nonumber\\
+&(1-\delta_{n,0})\sum_{r=1}^{n-1}{n \choose r}{\mathcal M}_{\mu+r}{\mathcal M}_{\nu+n-r},~n=0,1,\ldots
\end{align}
which for $n=0,2$ may be explicitly written as
\bes
\label{eq:moments}
\begin{align}
\label{eq:moment1}
\lambda ({\mathcal M}_{\mu+1}{\mathcal M}_\nu+{\mathcal M}_\mu{\mathcal M}_{\nu+1})&=
(1+2\lambda){\mathcal M}_\mu{\mathcal M}_\nu, \\
 \lambda ({\mathcal M}_{\mu+2}{\mathcal M}_\nu+{\mathcal M}_\mu{\mathcal M}_{\nu+2})&=
2{\mathcal M}_{\mu+1}{\mathcal M}_{\nu+1}\nonumber\\
&+(1+2\lambda){\mathcal M}_\mu{\mathcal M}_\nu.
 \label{eq:2nd-mom-eq}
\end{align}
\ees

Given the small mass behavior of $N(m)$ as in \eref{eq:smallmass}, the dependence of the  $\alpha$-th moment of mass on 
the cutoff mass $M$ may be determined as: 
\be
\label{eq:integral}
\mathcal{M}_\alpha\sim a_s M^{-\eta_s}\int^M dm \;m^{\alpha-\tau_s},
\ee
where by $x\sim y$, we mean that  $x/y=O(M^0)$ when $\lambda \to 0$. There is 
no divergence at small masses as the integral is cut off at the smallest mass $m_0=1$.
Thus, we obtain
\begin{align}
\label{eq:moment}
\mathcal{M}_\alpha\sim
	\begin{cases}
	M^{-\eta_s} \ln M, & \alpha=\tau_s-1,\\
	M^{-\eta_s+\max(\alpha+1-\tau_s,0)}, & \alpha\neq\tau_s-1.
	\end{cases}
\end{align}

We first derive  upper and lower bounds for the exponent $\tau_s$. We first show that $\tau_s < \nu+2$. Assume
that $\tau_s > \nu+2$.  We immediately obtain  from \eref{eq:moment} that $\mathcal{M}_\mu\sim \mathcal{M}_{\mu+1}\sim \mathcal{M}_\nu\sim \mathcal{M}_{\nu+1}\sim M^{-\eta_s}$. In this case, \eref{eq:moment1} simplifies to
$\lambda M^{-2 \eta_s} \sim M^{-2 \eta_s}$ or equivalently $\lambda\sim O(1)$. But, $\lambda$ is a parameter which tends to 0, hence, we arrive at a contradiction. Hence, $\tau_s \le \nu+2$. We now  show that $\tau_s\neq \nu+2$. Assume
$\tau_s=\nu+2$. We immediately obtain  from \eref{eq:moment} that $\mathcal{M}_\mu\sim \mathcal{M}_{\mu+1}\sim \mathcal{M}_\nu\sim M^{-\eta_s}$, and $\mathcal{M}_{\nu+1}\sim M^{-\eta_s} \ln M$. It is then straightforward to obtain from
\eref{eq:moment1} that $\lambda \sim 1/\ln M$. Knowing that $\mathcal{M}_{\nu+2}\sim M^{1-\eta_s}$, \eref{eq:2nd-mom-eq}
simplifies to $\lambda M \sim \ln M$ or $\lambda \sim M^{-1} \ln M$,  in contradiction with the earlier result  
$\lambda \sim 1/\ln M$. Hence,
we conclude that $\tau_s < \nu+2$.

We now show that $\tau_s > \mu+1$. Suppose $\tau_s<\mu+1$. Then, from Eq.~\eqref{eq:moment}, $\mathcal{M}_{\mu+n}\sim M\mathcal{M}_\mu$ and $\mathcal{M}_{\nu+n}\sim M\mathcal{M}_\nu$ for $n\geq 0$. 
In this case,  \eref{eq:2nd-mom-eq} simplifies to 
$\lambda M^2\mathcal{M}_\mu\mathcal{M}_\nu\sim M^2\mathcal{M}_\mu\mathcal{M}_\nu$
or $\lambda \sim O(1)$. But, $\lambda$ is a parameter which tends to 0, hence we arrive at a contradiction. Hence,
$\tau_s \geq \mu+1$. We now show that $\tau_s \neq \mu+1$. In this case, from \eref{eq:moment}, it follows
that $\mathcal{M}_\mu \sim M^{-\eta_s} \ln M$,   $\mathcal{M}_{\mu+n}\sim M\mathcal{M}_\mu/\ln M$, and 
$\mathcal{M}_{\nu+n}\sim M\mathcal{M}_\nu$ for $n\geq 0$. It is straightforward to show that substituting into 
\eref{eq:moment1} gives $\lambda \sim M^{-1}$, while substituting into \eref{eq:2nd-mom-eq} gives
$\lambda \sim 1/\ln M$, leading to a contradiction. We thus obtain $\tau_s > \mu+1$.
Combining the two bounds,
\be
\label{eq:bounds}
\mu+1<\tau_s<  \nu+2.
\ee

The equations for moments [see Eq.~\eqref{eq:moments}] may be further simplified
if  only the order of magnitude of the different terms is considered. Consider \eref{eq:moment1}.
We will argue that the left hand side of \eref{eq:moment1} is dominated by the
second term. Let $r={\mathcal M}_\mu {\mathcal M}_{\nu+1}/({\mathcal M}_{\mu+1}
{\mathcal M}_{\nu})$. 
If the integral \eqref{eq:integral} determining $ {\mathcal M}_{\nu}$
does not diverge, then neither will the integral for $ {\mathcal M}_{\mu}$ diverge, implying that 
${\mathcal M}_{\nu} \sim {\mathcal M}_{\mu}$. Then, 
$r \sim {\mathcal M}_{\nu+1}/{\mathcal M}_{\mu+1}$.  Since $\nu\geq \mu$, clearly $r \sim O(M^x)$ 
where $x\geq 0$.
On the other hand, if the integral for 
$ {\mathcal M}_{\nu}$ diverges,
then ${\mathcal M}_{\nu+1}\sim  M {\mathcal M}_{\nu}$. Then,
$r \sim {\mathcal M}_\mu M/{\mathcal M}_{\mu+1}$. Since 
${\mathcal M}_{\mu+1} /{\mathcal M}_{\mu}$ can diverge utmost as $M$,
we again obtain  $r \sim O(M^x)$ 
where $x\geq 0$.  \Eref{eq:moment1} then reduces to
$\lambda {\mathcal M}_{\nu+1} \sim {\mathcal M}_{\nu}$, or, equivalently,
$\lambda  \sim  {\mathcal M}_{\nu}/ {\mathcal M}_{\nu+1}$.

The same reasoning may be used to argue that the
left hand side of \eref{eq:2nd-mom-eq} is dominated by the second term. The
left hand side is then  $\lambda {\mathcal M}_{\mu}  
{\mathcal M}_{\nu+2}\sim 
{\mathcal M}_{\mu}  {\mathcal M}_{\nu} 
{\mathcal M}_{\nu+2}/{\mathcal M}_{\nu+1}$, where
we substituted for $\lambda$.
Since $\tau_s < \nu+2$ [see \eref{eq:bounds}], $ {\mathcal
M}_{\nu+2} /{\mathcal M}_{\nu+1} \sim M$, and the left hand side simplifies to
$ M {\mathcal M}_{\mu}  {\mathcal M}_{\nu} $.
The right hand side of \eref{eq:2nd-mom-eq} has to be then dominated by 
$ 2 {\mathcal M}_{\mu+1}  {\mathcal M}_{\nu+1}$.
The equations for moments [see Eq.~\eqref{eq:moments}] may then be rewritten
as
\begin{subequations}
\label{eq:momentsnew}
\bea
\frac{{\mathcal M}_{\nu}}{ {\mathcal M}_{\nu+1}}
&\sim& \lambda,
\label{eq:m0} \\
{\mathcal M}_{1}&\sim& 1,
\label{eq:m1} \\
{\mathcal M}_{\mu+1}
{\mathcal M}_{\nu+1}& \sim & M {\mathcal M}_{\mu}
{\mathcal M}_{\nu}, \label{eq:m2}
\eea
\end{subequations}
where \eref{eq:m1} follows from conservation of mass.

We can now derive $\tau_s$, $\eta_s$, and $y$ in terms of the known parameters. We have already shown that $\mu+1\leq\tau_s<\nu+2$  [see \eref{eq:bounds}]. For this range of $\tau_s$, and applying Eq.~\eqref{eq:moment}, we obtain
\begin{subequations}
\label{eq:moments_behavior}
\bea
{\mathcal M}_{\mu}&\sim & M^{-\eta_s},\\
{\mathcal M}_{\mu+1}&\sim & M^{-\eta_s+\max(\mu+2-\tau_s,0)},\\
{\mathcal M}_\nu &\sim & M^{-\eta_s+\max(\nu+1-\tau_s,0)},\\
{\mathcal M}_{\nu+1} &\sim &  M^{-\eta_s+\nu+2-\tau_s}.
\eea
\end{subequations}
Substituting  \eref{eq:moments_behavior} into \eref{eq:momentsnew}, we obtain
\begin{subequations}
\label{eq:exponent-identities} 
\begin{align}
\label{eq:y}
&\frac{1}{y}=\nu+2-\tau_s - \max(\nu+1-\tau_s,0),\\
\label{eq:etas}
&\eta_s = \max(2-\tau_s,0),\\
\label{eq:taus}
&\max(\nu+1-\tau_s,0) = \max(\mu+2-\tau_s,0)+\nu+1-\tau_s.
\end{align}
\end{subequations}

To make further progress, we consider different regimes of $\tau_s$. 
Consider first $\tau_s<\nu+1$. \Eref{eq:exponent-identities} 
implies that 
\begin{subequations}
\begin{align}
&y =1,\label{eq:y-non-local}\\
&\eta_s = \max(2-\tau_s,0), \quad\theta>1, \label{eq:eta-nonlocal}\\
&\mu+2\leq \tau_s < \nu+1,\label{eq:35c}
\end{align}
\end{subequations}
where we obtained the constraint on $\theta$ from requiring that a non-zero interval should exist
for the inequality satisfied by  $\tau_s$ in \eref{eq:35c}.   
Note that the values of $\tau_s$ and $\eta_s$ cannot be determined using moment analysis alone.

\begin{figure}
\includegraphics[width=\columnwidth]{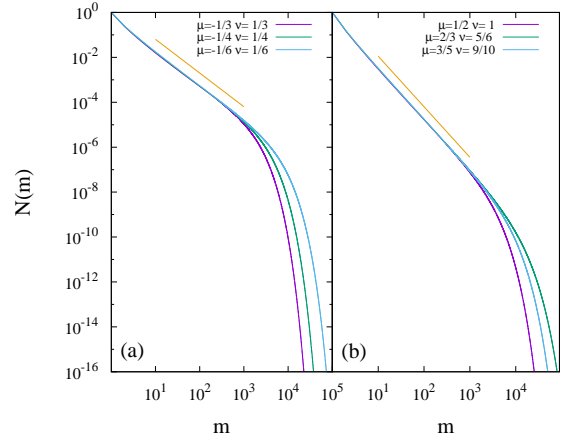}
\caption{\label{fig04}The steady mass distribution $N(m)$ for kernels with fixed $\beta$ and different $\theta<1$. (a) $\beta=0$ for $\theta=2/3,1/2, 1/3$, and 
(b) $\beta=3/2$ for $\theta= 1/2, 1/6, 3/10$. The solid lines are power laws with exponents $(3+\beta)/2$, as derived in
Eq.~(\ref{eq:tau-local}). The data are for $\lambda=0.01$}.
\end{figure}

Consider now the second case: $\tau_s>\nu+1$. In this regime, 
\Eref{eq:y} implies that  $y^{-1}=\nu+2-\tau_s$, while
\eref{eq:taus} reduces to
\be
\label{eq:constraint}
\nu+1-\tau_s+\max(\mu+2-\tau_s,0)=0.
\ee 
If $\tau_s\geq \mu+2$, then \eref{eq:constraint} implies that $\tau_s=\nu+1$ but we had 
assumed that $\tau_s>\nu+1$. Therefore, we conclude that $\tau_s<\mu+2$. This, in conjunction with 
the assumption $\tau_s>\nu+1$, implies that $\theta=\nu-\mu<1$, i.e., the kernel is local. 
We immediately obtain from \eref{eq:constraint} that $\tau_s=(3+\beta)/2$. Knowing $\tau_s$ allows to derive all the exponents. To summarize,
\begin{subequations}
\bea
\tau_s &=& \frac{3+\beta}{2}, \label{eq:tau-local}\\
\eta_s &=& \max\left[\frac{1-\beta}{2},0 \right], \quad \theta<1 \label{eq:eta-local},\\
y &=&\frac{2}{\theta+1}.  \label{eq:y-local}
\eea
\end{subequations}
We now verify numerically that the correctness of
Eq.~(\ref{eq:tau-local}) for $\theta<1$. 
In Fig.~\ref{fig04}, we show the variation of $N(m)$ with $m$ for two
different values of $\beta$, and varying $\theta<1$. 
The data for $N(m)$ for small masses are independent of $\theta$, and
are consistent with a power law with exponent given by 
Eq.~(\ref{eq:tau-local}).

Thus when the kernel is  local, all exponents describing the small mass behavior of the 
mass distribution can be obtained using moment analysis, unlike  the case when the kernel is 
non-local. However, the analysis of singularities will enable us to determine the unknown exponents.

We now study the case when $\theta=\nu-\mu=1$, the boundary between
the kernel being local or  non-local.  For this special case, we expect that the power laws will be modified by additional logarithmic 
corrections~\cite{Connaughton17}. We assume 
the following form for $N(m)$:
\be
N(m) \sim \frac{ (\ln m)^{-x} (\ln M)^{-z}}{m^{\nu+1}
M^{\eta_s}},~m \ll M,~\theta=1,
\label{eq:borderline}
\ee
where the cutoff mass scale $M$ could have a  logarithmic dependence on 
$\lambda$, and $x$ and $z$ are new exponents that characterize the logarithmic
corrections.  The choice of $\tau_s=\nu+1$ is motivated from $\theta\to 1$ behavior
of  \eref{eq:tau-local}.
It is then straightforward to obtain 
${\mathcal M}_\mu \sim M^{-\eta_s} (\ln M)^{-z}$, 
${\mathcal M}_{\mu+1} \sim {\mathcal M}_\nu
\sim M^{-\eta_s} (\ln M)^{-z+\max(0,1-x)}$, and 
${\mathcal M}_{\nu+1} \sim M^{1-\eta_s} (\ln M)^{-x-z}$. 
For $\nu=\mu+1$, \eref{eq:m2} reduces to $
{\mathcal M}_{\nu+1} \sim  M {\mathcal M}_{\mu}
$. Substituting for the moments, we immediately obtain
\be
x=0.
\ee
For this choice of $x$, \eref{eq:m0} immediately yields $\lambda \sim M^{-1} \ln M$ or
\be
M \sim \frac{-\ln \lambda}{\lambda}, \quad \theta=1.
\ee
Substituting for the different moments into \eref{eq:m1}, it is straightforward to derive
\bea
\eta&=& \max (1-\nu,0),\quad \theta=1,\\
z&=& \delta_{\nu,1}, ~\quad \theta=1.
\eea

\section{Singularity analysis \label{sec:singularity}}

In this section, we analyze the equation [see \eref{eq:gen-fun}] satisfied by the generating functions 
$F_\mu(x)$ and $F_\nu(x)$, based on their singular behavior.
This will allow us to
determine the exponents $\tau_\ell$ and 
$\eta_\ell$ [see \eref{eq:exponential} for definition]. This, in turn, will allow us to
determine the exponents $\tau_s$ and $\eta_s$ characterizing the small mass behavior of $N(m)$
for non-local kernels. 

Let the singularity of $F_{\mu}(x)$ closest to the origin be denoted by
$x_c$. Comparing with Eq.~\eqref{eq:exponential}, we immediately obtain
\be
M=\frac{1}{\ln x_c}.
\label{eq:Mx0}
\ee
Consider $x=x_c-\epsilon$, $\epsilon \to 0$. If the large behavior of $N(m)$ is as in
Eq.~\eqref{eq:exponential}, then the 
leading singular behavior of the
generating functions  $F_\nu$ and $F_\mu$ close to the singular point is
proportional to  $\epsilon^{\tau-\nu-1}$ and
$\epsilon^{\tau-\mu-1}$, respectively. Depending on the value of $\tau$, $F_\nu(x_c)$
or $F_\mu(x_c)$ may diverge or tend to a constant as $\epsilon \to 0$. 

Expressing $F_\nu(x)$ in terms of $F_\mu(x)$ from \eref{eq:gen-fun}, we obtain
\bea
\label{eq:gen-fun-new}
F_\nu(x)=\frac{(1+\lambda)\mathcal{M}_\nu F_\mu (x)-x(1+2\lambda)\mathcal{M}_\mu\mathcal{M}_\nu}{F_\mu(x)-(1+\lambda)\mathcal{M}_\mu}.
\eea
We now claim that $F_\mu(x_c) = (1+\lambda){\mathcal M}_\mu$. Suppose this were not the case and
$F_\mu(x_c) \neq (1+\lambda){\mathcal M}_\mu$. Then, the denominator in \eref{eq:gen-fun-new} may be
set to a constant when expanding about $x_c$, and it follows that $F_\nu(x)$ has the same singular
behavior as $F_\mu(x)$  near $x=x_c$. 
This implies that $\mu=\nu$. When $\mu=\nu$, we have determined the generating function $F_\mu(x)$
exactly (see Sec.~\ref{sec:exact}), and 
it is easily seen from \eref{eq:fmu} that $F_\mu(x_c)= (1+\lambda){\mathcal M}_\mu$. This contradicts
our initial assumption that  $F_\mu(x_c) \neq (1+\lambda){\mathcal M}_\mu$.
When $\mu\neq\nu$,
$F_\mu(x)$ and $F_\nu(x)$ should have different singular singular behavior near $x=x_c$, again leading
to a contradiction. We therefore
conclude that
\be
F_\mu(x_c)=(1+\lambda){\mathcal M}_\mu.
\label{eq:eq40}
\ee

Expanding the generating functions about $x=x_c$, we obtain
\begin{subequations}
\label{eq:fsingularIII}
\begin{align}
&F_\mu(x_c\!-\!\epsilon) =   (1+\lambda){\mathcal M}_\mu- \epsilon^{\tau_\ell-\mu-1} R_1(\epsilon) -
\epsilon R_2(\epsilon),\\
&F_\nu(x_c\!-\!\epsilon) =  \epsilon^{\tau_\ell-\nu-1} R_3(\epsilon) + R_4(\epsilon),
\end{align}
\end{subequations}
where $R_i$'s are regular in $\epsilon$, $R_1(0) \neq 0$, and 
$R_3(0) \neq 0$. Also, 
\be
\tau_\ell > \mu+1,
\label{eq:boundtauell}
\ee
so that \eref{eq:eq40} is satisfied.
We now examine the numerator of \eref{eq:gen-fun-new} when $x=x_c$. 
On simplifying by using \eref{eq:eq40}, it reduces to $(1+2\lambda){\mathcal M}_\mu {\mathcal M}_\nu
[1+\lambda^2-x_c+O(\lambda^3)]$. However, $x_c \sim 1+M^{-1}\sim 1+\lambda^{y}$ when $\lambda \to 0$. 
We have shown earlier  that $y<2$ for $\theta>0$ [see Eqs. (\ref{eq:y-non-local}) and (\ref{eq:y-local})].
Thus, the numerator of \eref{eq:gen-fun-new} is non-zero and equal to 
$-{\mathcal M}_\mu {\mathcal M}_\nu M^{-1}$,
when $x=x_c$, $\lambda \to 0$,  and $\theta>0$.
Substituting the expansions [\eref{eq:fsingularIII}] into \eref{eq:gen-fun-new}
we obtain
\be
\label{eq:sing-main}
\epsilon^{\tau_\ell-\nu-1} R_3(\epsilon) + R_4(\epsilon)=\frac{-{\mathcal M}_\mu {\mathcal M}_\nu M^{-1}}{- \epsilon^{\tau_\ell-\mu-1} R_1(\epsilon) +
\epsilon R_2(\epsilon)}.
\ee
Since $\tau_\ell > \mu+1$ [see \eref{eq:boundtauell}], the right hand side of \eref{eq:sing-main} diverges. This implies
that
\be
\tau_\ell <\nu+1,
\label{eq:boundtauell2}
\ee
and the left hand side of \eref{eq:sing-main} is dominated by the first term.
We can now compare the leading  singular behavior on  both sides of \eref{eq:sing-main}.
There are two possible cases: $0<\tau_\ell-\mu-1 <1$ and $0<\tau_\ell-\mu-1 >1$.
\subsection{$\tau_\ell-\mu-1 <1$}
\begin{figure}
\includegraphics[width=\columnwidth]{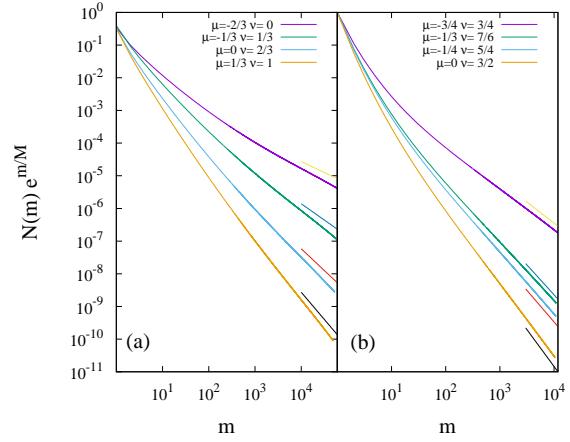}
\caption{
\label{fig05}The compensated steady mass distribution $N(m) e^{m/M}$ 
for kernels with fixed $\theta<2$ and different $\beta$. 
(a) $\theta=2/3$ for $\beta=-2/3,0,2/3,4/3$ and
(b) $\theta=3/2$ for $\beta= 0,5/6,1,3/2$. The solid lines are power
laws with exponents $(2+\beta)/2$, as derived in
Eq.~(\ref{eq:tau-l-local}). The data are for $\lambda=0.01$.}
\end{figure}

First consider the regime
$0<\tau_\ell-\mu-1 <1$.  The denominator of Eq.~\eqref{eq:sing-main}  is dominated by 
$- \epsilon^{\tau_\ell-\mu-1} R_1(\epsilon)$, and comparing the 
singular terms on both sides,
we obtain 
\be
\tau_\ell=\frac{\beta+2}{2}, \quad \theta<2\label{eq:tau-l-local},
\ee
where we obtain the constraint on $\theta$ from our assumption 
$0<\tau_\ell-\mu-1 <1$. 
We now verify numerically that the correctness of
Eq.~(\ref{eq:tau-l-local}) for $\theta<2$.
In Fig.~\ref{fig05}, we show the variation of $N(m)$ with $m$ for 
two values of $\theta$, one between zero and one and the
other between one and two, and varying $\beta$.
The data for compensated mass distribution for large masses 
are consistent with a power law with exponent given by
Eq.~(\ref{eq:tau-l-local}).

Comparing now the coefficients of the leading singular terms we obtain
\be
R_3(0) R_1(0)= {\mathcal M}_\mu {\mathcal M}_\nu M^{-1}, \quad \theta <2.
\ee
Once  $R_1(0)$, $R_3(0)$ and
$\tau_\ell$ are known, $m^\mu N(m)$ and $m^\nu N(m)$ may be obtained from $F_\mu(x)$
and $F_\nu(x)$ by doing inverse Laplace transforms. Thus
\bea
m^\mu N(m)&=& \frac{- R_1(0) x_c^{\tau_\ell-\mu -1} (\tau_\ell -\mu-1)}{x_c^m 
m^{\tau_\ell-\mu}\Gamma(2-\tau_\ell+\mu)}, \label{eq:ee}\\
m^\nu N(m)&=& \frac{R_3(0) x_c^{\tau_\ell-\nu -1} (\tau_\ell -\nu-1)}{x_c^m 
m^{\tau_\ell-\nu}\Gamma(2-\tau_\ell+\nu)}. \label{eq:ff}
\eea
where $\Gamma(x)$ is the Gamma function. Multiplying together Eqs.~\eqref{eq:ee}, and \eqref{eq:ff},
setting $\tau_\ell= (2+\beta)/2$ [see \eref{eq:tau-l-local}],
and using the property
\be
\Gamma(x) \Gamma(1-x)= \frac{\pi}{\sin (\pi x)},
\ee
we obtain
\be
N(m) \simeq \sqrt{\frac{{\mathcal M}_\mu {\mathcal M}_\nu \theta \sin\frac{\pi \theta}{2}}{2 \pi M}} 
\frac{e^{-m/M}}{m^{(2+\beta)/2}},~ m \gg M,
\ee
for $0<\theta<2$.
The prefactor depends on ${\mathcal M}_\mu$ and $ {\mathcal M}_\nu$, which are determined by the 
behavior of $N(m)$ at small masses. Their dependence on the cutoff $M$ [see \eref{eq:moments_behavior}] 
will determine $\eta_\ell$:
\be
\eta_\ell=\frac{1}{2}+\eta_s - \frac{1}{2} \max(\nu+1-\tau_s,0).
\label{eq:99}
\ee

Knowing $\tau_\ell=(2+\beta)/2$ [see \eref{eq:tau-l-local}], the relation
$\tau_s+ \eta_s=\tau_\ell + \eta_\ell$ [see \eref{eq:expequality}] reduces to
\be
\tau_s= \frac{3+\beta}{2}- \frac{1}{2} \max(\nu+1-\tau_s,0).
\label{eq:100}
\ee
We consider the two cases $\nu+1-\tau_s<0$ and $\nu+1-\tau_s>0$ separately.

{\it Case I: $\nu+1-\tau_s<0$}.
In this case, \eref{eq:100} immediately gives $\tau_s=(3+\beta)/2$. To satisfy the inequality $\nu+1-\tau_s<0$, we
require that $\theta<1$. This result for $\tau_s$ is consistent with what we derived earlier for the local kernel using
moment analysis [see \eref{eq:tau-local}]. Knowing $\tau_s$ and $\eta_s=\max[(1-\beta)/2,0] $ 
[see \eref{eq:eta-local}], we obtain from \eref{eq:99}
\be
\eta_\ell= \frac{1}{2} +  \max\left[\frac{1-\beta}{2},0 \right], \quad \theta<1.
\ee

{\it Case II: $\nu+1-\tau_s>0$}. 
In this case, \eref{eq:100} immediately gives 
\be
\tau_s=2+\mu=\frac{4+\beta-\theta}{2},~ \quad 1<\theta<2, \label{eq:101}
\ee
where the constraint on $\theta$ is obtained from the 
inequality $\nu+1-\tau_s<0$.  This result for $\tau_s$ is
consistent with the inequality derived for $\tau_s$ using moment analysis [see \eref{eq:35c}],
Knowing $\tau_s$, $\eta_s$, and $\eta_\ell$ may be derived from the Eqs.~(\ref{eq:eta-nonlocal}) and (\ref{eq:99})
to be
\bea
\eta_s&=& \max\left[-\mu, 0 \right], \quad 1<\theta<2,\\
\eta_\ell&=& \frac{2-\theta}{2} +  \max\left[-\mu, 0 \right], \quad 1<\theta<2.
\eea
For $\tau_s=\mu+2$, then there is the possibility of logarithmic corrections.

Thus, we have derived all the exponents characterizing both the small and large mass behavior of $N(m)$
when $\theta<2$.

\subsection{$\tau_\ell-\mu-1 >1$}
Consider now the second case when $\tau_\ell-\mu-1 >1$. The denominator of Eq.~(\eqref{eq:sing-main}) 
is dominated by $\epsilon R_2(\epsilon)$.   Again comparing the singular terms on both sides of  
Eq.~\eqref{eq:sing-main}, we obtain
\be
\tau_\ell=\nu, \quad \theta>2, 
\label{eq:tau-l-addition}
\ee 
where we obtain the constraint in $\theta$ from our assumption $\tau_\ell-\mu-1 >1$ .
Comparing the coefficients of the leading singular terms we obtain
\be
R_2(0) R_3(0)= {\mathcal M}_\mu {\mathcal M}_\nu M^{-1}, \quad \theta>2.
\label{eq:53}
\ee
It is easy to see that $R_2(0) = F_{\mu+1}(x_c)$.
Doing an inverse Laplace transform, we obtain
\be
N(m) \simeq \frac{m^{-\nu}}{MF_{\mu+1}(x_c)} e^{-m/M}, ~m \gg M,~\theta>2.
\label{eq:61}
\ee

The dependence of $R_2(0) = F_{\mu+1}(x_c)$ on $M$ may be determined as follows.
The integral for $F_{\mu+1}(x_c)$ has two power laws: 
\be
F_{\mu+1}(x_c) \sim \int^M dm \frac{m^{\mu+1}}{M^{\eta_s} m^{\tau_s}}+
 \int_M^\infty dm \frac{m^{\mu+1}}{M^{\eta_\ell} m^{\nu}}
\ee
Using the bound Eq.~\eqref{eq:35c},
it is straightforward to argue that
to leading order $F_{\mu+1}(x_c) \sim M^{-\min(\eta_s,\eta_\ell+\theta-2)}$. 
Substituting $R_3(0) \sim M^{-\eta_\ell} $
and $R_2(0) \sim M^{-\min(\eta_s,\eta_\ell+\theta-2)}$ into
Eq.~\eqref{eq:53}, we immediately obtain
\be
\eta_\ell+\min(\eta_s,\eta_\ell+\theta-2) = 2\eta_s, \quad \theta>2.
\label{eq:102}
\ee
We consider the two cases $\eta_s<\eta_\ell+\theta-2$ and $\eta_s>\eta_\ell+\theta-2$ separately.

{\it Case I: $\eta_s<\eta_\ell+\theta-2$}.
From \eref{eq:102}, we obtain
\be
\eta_\ell = \eta_s, \quad \theta>2,
\ee
where the constraint on $\theta$ is obtained from the assumption that $\eta_s<\eta_\ell+\theta-2$.
\Eref{eq:expequality} then yields $\tau_s=\tau_\ell$. Therefore,
Eqs.~(\ref{eq:tau-l-addition}) and (\ref{eq:eta-nonlocal}) imply  that 
\bea
\tau_s &=& \nu, \label{eq:nu} \\
\eta_s &=& \max(2-\nu,0),\quad \theta>2,\\
\eta_\ell &=& \max(2-\nu,0).
\eea

{\it Case II: $\eta_s>\eta_\ell+\theta-2$}: 
From \eref{eq:102}, we obtain
\be
\eta_\ell = \eta_s + 1 -\frac{\theta}{2}.
\ee
This solution in conjunction with our assumption that $\eta_s>\eta_\ell+\theta-2$ imply that
$\theta<2$. But, the solution \eref{eq:tau-l-addition} is valid only for $\nu>2$. Hence there
is no solution for this case. 
We note that the results for $\tau_\ell$ and $\eta_\ell$ 
coincide with those for the addition model when $\theta>2$ [see \eref{eq:addition-main}] .

\begin{figure}
\includegraphics[width=\columnwidth]{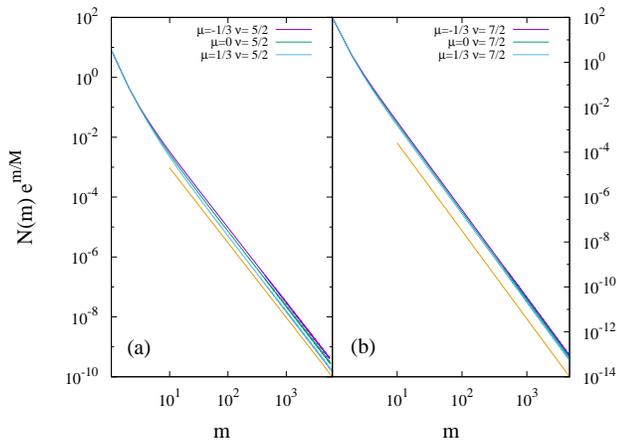}
\caption{Left: $\nu=1.5; \mu=-1$. Right: $\nu=2.5; \mu=-0.33$ . Left: does not scale as $\lambda^{\eta_{S}}$ 
\label{fig06}The compensated steady mass distribution $N(m) e^{m/M}$ 
for kernels with fixed $\nu$ and different $\theta>2$. 
(a) $\nu=5/2/$ for $\theta=17/6,5/2,13/6$ and
(b) $\nu=7/2/$ for $\theta=23/6,7/2,19/6$. The solid lines are power
laws with exponents $\nu$, as derived in
Eq.~(\ref{eq:nu}). The data are for $\lambda=0.01$.}
\end{figure}

We now verify numerically that the correctness of
Eq.~(\ref{eq:nu}) for $\theta>2$.
In Fig.~\ref{fig06}, we show the variation of $N(m)$ with $m$ for 
two values of $\nu$, for different values of  $\theta>2$.
The data for compensated mass distribution for large masses 
are consistent with a power law with exponent given by
Eq.~(\ref{eq:nu}).



\section{\label{sec:rings}Aggregation-fragmentation models for planetary rings}

\begin{figure}
\includegraphics[width=\columnwidth]{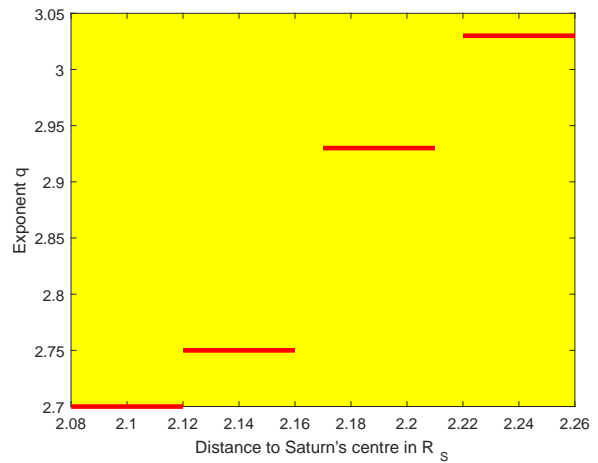}
\caption{\label{fig07} The scaling exponent $q$: $N(R)\sim R^{-q}$ extracted from the occultation
data obtained by the Voyager Radio Science Subsystem in~\cite{zebker1985}. The exponent is extracted 
separately for each of the four sub-regions of the A-ring. The size and the position of each region is indicated
by the horizontal red bars on the graph. All distances are measured in the units of Saturn's radius, $R_S$.} 
\end{figure}

 When background stars are occulted by the rings of Saturn, the properties of the scattered light depend on the particle size distribution of the rings. Observation of such occultations suggests a power law distribution of particle sizes. The exponent describing the distribution of particle sizes near the outer edge of the $A$-ring of Saturn extracted from Cassini data in Ref.~\cite{becker2015} varies between $3.5$
and $2.8$ depending on the star occulted.  Simple particle models of coalescence and shattering have been proposed to explain the power law scaling.  Model-based extractions~\cite{zebker1985} of the scaling exponent suggest an increase in steepness of the distribution with distance to Saturn (see 
Fig. \ref{fig07}). The graph shows an almost linear increase of the scaling exponent with 
distance (see Ref.~\cite{Cuzzi10} for an overview and Refs.~\cite{brilliantov2009,brilliantov2015size} for recent theoretical work).  In this section, we discuss the implications of our results for these  efforts. 
In the simplest model  it is supposed that binary collisions dominate the dynamics and the collision kernel, $K(m_1, m_2)$ depends on the velocity distribution of particles within the rings. The colliding particles coalesce into a single particle of mass $m_1+m_2$ with probability $1-p$ and shatter with probability $p$ to create `dust' - $\frac{m_1+m_2}{m_0}$ particles of the smallest mass $m_0$
present in the system. 
For example, the collision kernel obtained under the assumption of
energy equipartition of particles constituting Saturn's ring is
\begin{equation}
\label{intro_kerE}
K^{(E)}(m_1,m_2)=C\left|m_1^{-\frac{1}{2}}+m_2^{-\frac{1}{2}}\right| \left(m_1^{\frac{1}{3}}+m_2^{\frac{1}{3}}\right)^2 ,
\end{equation}
where $C$ is a positive constant. 
The first mass-dependent factor accounts for the relative  particle velocity and the second - for the area of collisional cross  section.
The homogeneity degree of the kernel (\ref{intro_kerE}) is $\beta=1/6$. The rings of Saturn are a non-equilibrium system, 
therefore, the equipartition of energy does not follow from any general principles. As is suggested in 
\cite{brilliantov2015size}, the whole range of velocity distributions from equipartition of energy to mass-independent root mean square
velocity might be present across the different subrings. The latter extreme leads to the following kernel:
\begin{eqnarray}\label{intro_kerV}
K^{(V_{rms})}(m_1,m_2)=C \left(m_1^{\frac{1}{3}}+m_2^{\frac{1}{3}}\right)^2,
\end{eqnarray}
which is just proportional to the geometrical cross section. This kernel is
also homogeneous with $\beta=2/3$. The kernels  (\ref{intro_kerE}) and  (\ref{intro_kerV}) belong to the class of homogeneous kernels described in the Introduction - their asymptotic behavior is captured by exponents $\nu$ and $\mu$.  Studying these kernels when $m_2 \gg m_1$ gives $\nu=2/3$ and $\mu=-1/2$ for the kernel (\ref{intro_kerE})
and $\nu=2/3$ and $\mu=0$ for the kernel (\ref{intro_kerV}). Therefore, $\theta=7/6$ for the energy equipartition kernel (\ref{intro_kerE}) and $\theta=2/3$ for the constant
root mean square velocity kernel (\ref{intro_kerV}).

Kernel (\ref{intro_kerV}) is local and the corresponding distribution of particle masses scales as $m^{-11/6}$.
This implies that the distribution of particle $radii$ is given by $ R^2 N(R^3)\sim R^{-7/2}$. The exponent
$7/2$ is consistent with the upper range of scaling exponents describing the distribution of constituent
sizes in Saturn's rings~\cite{jerousek2016small}. 
On the other hand, the energy equipartition kernel (\ref{intro_kerE}) is not local: the substitution of (\ref{intro_KZ}) with $\beta=1/6$
into the corresponding formula for the flux will lead to a divergence in the limit of small shattering probability.
Correspondingly, a conclusion of \cite{brilliantov2009,brilliantov2015size} that the mass distribution of particles in the coalescence-shattering model with the kernel (\ref{intro_kerE}) scales as $m^{-19/12}$ (equivalently, the distribution of particle sizes scales as $R^{-11/4}$) is probably incorrect  \footnote{The `experimental' curve plotted in Fig. $3$ from \cite{brilliantov2015size} does not
show  raw `data from Voyager RSS'. It was obtained in \cite{zebker1985} using model-based analysis of Voyager occultation data based on a
number of assumed
model parameters such as the particle's density. Moreover, the curve does not pertain to the whole A-ring, but just one of its sub-regions (A$2.14$) where the inferred exponent happens to be $2.75$.}
and there is no constant-flux scaling in this case.
According to the analysis of Secs. \ref{sec:moments} and \ref{sec:singularity} in the regime of weak non-locality ($1<\theta<2$), the correct answer for the mass distribution is
\begin{eqnarray}\label{intro_NLE} 
N(m)\sim m^{-\left(\frac{\beta+3}{2}-\frac{\theta-1}{2}\right)}.
\end{eqnarray}
Weak non-locality results in $\theta$-dependent correction to the Kolmogorov-Zakharov exponent.
For the kernel (\ref{intro_kerE}), this answer means that $N(m)\sim m^{-3/2}$, or the distribution
of particle sizes scales as $R^{-10/4}$. 

Of course, the exponents $10/4$ and $11/4$ are  indistinguishable from the observational point of view given that (i) there are no precise measurements of the spectral exponents for Saturn's rings (ii) there is no way to fix the parameters of the collision kernel from the existing data on the statistics of particles constituting the rings.   
By looking at a range of reasonable distributions of velocities of particles in the ring and solving the corresponding Smoluchowski equations the authors of~\cite{brilliantov2015size}  found the range of
scaling exponents describing the distribution of particle sizes to be $[2.75, 3.5]$, which agrees
with the numbers accepted by planetary scientists, see e. g. \cite{becker2015,jerousek2016small,colwell2018particle}. 
Accounting for the non-locality of the kernel (\ref{intro_kerE}), which corresponds to the left
boundary of the range, the interval should change to $[2.5, 3.5]$. This is neither here nor there,
as the observational knowledge of the exponents is not good enough to distinguish between these
predictions. Our conclusions are therefore of a more qualitative nature: the distribution of particles for
coalescence-shattering models with homogeneous kernels possessing a well-defined locality
exponent is indeed universal in the limit of small shattering probabilities.  However, the universality classes are
labeled by both the homogeneity degree $\beta$ {\em and} the locality exponent $\theta$ rather than by $\beta$ alone as suggested in  \cite{brilliantov2009,brilliantov2015size}. In the local
regime, $\theta<1$, the dependence of the mass distribution on $\theta$ disappears and we are left
with the constant flux distributions (\ref{intro_KZ}). For $\theta>1$ the mass distribution scales
with an exponent, which depends on both $\theta$ and $\beta$. We note that we only describe 
the universality classes of scaling $exponents$.
The $amplitude$ of $N(m)$ for non-local kernels is non-universal and depends explicitly on the
position of sources and sinks in mass space.

\section{Conclusion \label{sec:conclusion}}

In this paper, we determined the steady state mass distribution for a system of particles that on undergoing two-body
collisions either coalesce into a single particle or fragment into dust (particles of the smallest mass). The total mass is conserved 
by the dynamics. Fragmentation acts as a source of  particles of small mass while coagulation depletes smaller particles and creates
particles of larger mass.  We considered a class of homogeneous collision
kernels modeled by $K(m_1,m_2) = m_1^\mu m_2^\nu + m_1^\nu m_2^\mu$ with  $\nu \geq \mu$, characterized by 
the homogeneity exponent $\beta=\mu+\nu$ and non-locality exponent $\theta=\nu-\mu$. The results for the 
exponents characterizing the small and large mass 
distributions, obtained through a combination of moment analysis, singularity analysis, and exact solutions for special
cases, are summarized in Table~\ref{table1} for different $\beta$ and $\theta$.

The presence of a non-zero fragmentation rate $\lambda$  introduces a cutoff scale $M$
beyond which the mass distribution $N(m)$ crosses over from a power law 
behavior to an exponential decay with increasing mass $m$. 
Thus, a non-zero $\lambda$ is a useful regularization scheme by which 
instantaneous gelation is prevented for kernels that 
are gelling ($\mu+\nu>1$) and one may study the behavior as the 
regularization is removed by taking the limit $\lambda \to 0$. Here,
we find that the form of $N(m)$ depends only on whether the kernel is local
($\theta <1$) or non-local ($\theta \geq 1$) and
not on whether it is gelling or non-gelling.

We find two distinct non-local regimes corresponding to $1 < \theta < 2$ 
and $\theta  > 2$. When $\theta<1$, the distribution
is universal in the sense that the small mass behavior does not
depend on the source or sink. Thus, the limit $\lambda \to 0$ is well
defined. In the regime, $1 < \theta < 2$ the mass distribution
$N(m)$  depends on the sink scale $M$ but
is independent of the source scale, $m_0$. In the
regime $\theta  > 2$, $N(m)$ depends on both source and sink.
Logarithmic corrections are found at the boundaries between
regimes. These are similar to the two non-local
regimes that we found for the non-conserved model
driven by input of particles at small masses and collision-dependent evaporation~\cite{Connaughton17}.  The logarithmic corrections are  also analogous to the 
correction proposed by Kraichnan~\cite{kraichnan_inertial-range_1971} to account for the marginal
non-locality of the enstrophy cascade in two-dimensional fluid
turbulence. 

As we saw in Sec. \ref{sec:rings}, the study of Saturn rings provides examples of both local and non-local kernels.   
Clearly, there are many open questions relating to the size distribution of ring particles.  
Can we determine the kernels describing particle collisions in different parts of Saturn's rings  so that more quantitative predictions of particle size distributions can be made? In particular,
can the dependence in Fig.~\ref{fig07} be confirmed theoretically? 
Can one predict regions within the rings of Saturn, where the scaling is of Kolmogorov-Zakharov or constant flux type? 
Are there regions dominated by weak non-locality or regions correctly described
by the addition model? 
Answering these questions could open up interesting new avenues of research into coalescence-fragmentation models.

 Our results also have implications for the addition model in which clusters grow only through
reactions with the monomer. The addition model has been studied both as a model for island 
diffusion though desorption and adsorption of monomers as well as a solvable approximation
for more complicated collision kernels~\cite{Hendriks1984,brilliantov1991,laurencot1999,ball_instantaneous_2011,blackman1994coagulation,chavez1997some}. 
While the time-dependent as well as steady state solutions have been determined for
the addition model, in the latter case,  it is not clear when this approximation of restricting collisions only with monomers reproduces the
same result as the original kernel. The results of this paper show that
for $\theta >2$ the exponents characterizing the mass distribution for the general kernel
coincide with that for addition model for the same $\theta$. Thus, we conclude that 
the addition model is a good description of systems with $\theta >2$. 

In this paper, we studied the steady state but not the
dynamics leading to it. This question is important to consider. 
Even in the local case, $\theta<1$, the
dynamics leading to the steady state must be very different
for gelling ($\beta>1$) and non-gelling ($\beta<1$) kernels.
Furthermore, in the non-local case, evidence from
closely related models~\cite{ball_collective_2012,ball_instantaneous_2011} 
suggests that the steady
state could become unstable for $\lambda \to 0$. Such an instability
would result in persistent oscillatory kinetics. Indeed, such oscillations have been seen in a recent paper~\cite{matveev2017oscillations}. 
This would have interesting
consequences for the mass distribution in Saturn rings which could be 
experimentally verifiable.

In other models of aggregation and fragmentation, where fragmentation occurs spontaneously and not due to
a collision, an interesting phase transition occurs when the fragmentation is limited to a finite mass chipping
off to a 
neighbor~\cite{MKB1,MKB2,krapivsky1996transitional,rajesh2001exact,rajesh2002effect,rajesh2002aggregate}.
This model
undergoes a nonequilibrium phase transition from a phase in
characterized by an exponential mass distribution to a
phase characterized by power law mass distribution in the presence of a condensate. The condensate is
one single mass which carries a finite fraction of the total mass. It would be interesting to see whether the
model considered in the paper exhibits a similar transition in some parameter regimes.

In this paper, we have assumed that the system is well mixed, and hence it was possible to 
ignore spatial variations in the densities. Also, the effects of stochasticity were completely
ignored. Introducing stochasticity, even at the level of zero dimensions, can give rise to new
phenomenology like an absorbing-active phase transition in the $\lambda$-density plane.
This is because, if total mass is small enough, then the system has a finite probability of
getting stuck in an absorbing state where all particles have coalesced into one particle.
Including spatial variation would make the problem even richer. This is a promising area for 
future study.

\begin{acknowledgments}
C.C. and A.D. acknowledge funding from the EPSRC (Grant No. EP/M003620/1). A.D. also thanks the International Centre for Theoretical Sciences (ICTS) for support during a visit for participating in the program -Bangalore School On Statistical Physics - VII (Code No. ICTS/Prog-bssp/2016/07).
\end{acknowledgments}

%

\end{document}